\documentstyle[preprint,aps,eqsecnum,epsf,floats]{revtex}
\tighten

\begin{document}

{\tighten
\preprint{\vbox{\hbox{DUKE-94-77}
\hbox{CMU--HEP95--10}
\hbox{DOE--ER/40682--100}}}

\title{
Kaon-Nucleon Couplings for Weak Decays of Hypernuclei}
\author{Martin  J. Savage}
\address{
Department of Physics, Carnegie Mellon University, Pittsburgh, PA 
15213
\\ {\tt savage@thepub.phys.cmu.edu}}
\author{Roxanne P. Springer}
\address{Duke University Department of Physics, Durham, NC 27708
\\ {\tt rps@phy.duke.edu}}
\maketitle 

\begin{abstract}

We investigate the weak kaon-nucleon (NNK)  
S-wave and P-wave interactions using heavy baryon chiral 
perturbation theory.  
The leading 1-loop SU(3) breaking contributions to the 
$ppK$, $pnK$, and $nnK$ couplings are computed.
We find that they suppress all  
NNK amplitudes  by $30\%$ to $50\%$.   
The ratio of neutron-induced to proton-induced hypernuclear decay 
widths is sensitive to such reductions. 
It has been argued that the discrepancy between the 
predicted and observed P-wave amplitudes in $\Delta s=1$ hyperon 
decay 
results from an accidental cancellation between tree level 
amplitudes, 
and is not a fundamental problem for chiral perturbation theory.
Agreement between experimentally determined NNK P-wave amplitudes 
and  our  estimates  would support this explanation.

\end{abstract}
\pacs{13.75.Jz,21.80.+a,25.80.Pw,13.30.Eg}

\bigskip
\date{July 1995}
\vfill\eject

\section{Introduction}

Weak decays of hypernuclei provide a laboratory for 
investigating kaon-nucleon interactions. 
In free space, a hyperon such as $\Lambda$, $\Sigma$ or $\Xi$ 
will decay  predominantly through
a mesonic mode, e.g. $\Lambda\rightarrow N\pi$. 
However, when the hyperon is bound inside a nucleus with $A \sim 
12$ or 
larger, 
this decay mode is suppressed by Pauli-blocking of the final 
state 
nucleon (produced with $\vec{p}_N \sim 100 $ MeV/c, much less 
than 
the Fermi-momentum $\vec{p}_F \sim 280$ MeV/c).
A competing process that does not suffer significantly 
from Pauli-blocking is nonmesonic weak scattering, e.g.
$\Lambda N\rightarrow NN$, in which the final state nucleons have 
$\vec{p}_N \sim 400$ MeV/c.
These processes occur at low momentum scales where
QCD is in the nonperturbative regime, so the 
structure and decays of $\Lambda$, $\Sigma$, and $\Xi$ 
hypernuclei have been investigated using various phenomenological 
models,
e.g.
\cite{MG84,D86,OS8586,CHK83,HK86,N88,MDG88,C90,RMBJ92,D92,ITO94,MS94,MJ95}.
Available experimental observables for analyzing such systems 
include spin-averaged decay rates, 
the proton asymmetry from polarised hypernuclei,
and the ratio of neutron induced to proton induced 
decay widths.
The later proves especially difficult to describe using
hadronic models, implying that we do not yet have a complete 
understanding of the dynamics of these systems.

The weak scattering process receives 
contributions from long distance meson exchange diagrams 
and from short distance (compared to the scale of chiral symmetry 
breaking, 
$\Lambda_\chi \sim$ 1 GeV) four-baryon contact terms.  
The leading meson exchange graphs are the one-pion 
exchange (OPE) followed by 
one-kaon exchange (OKE), one-eta exchange (OEE) 
and two-pion exchange (TPE) and so forth. 
The OPE amplitudes can be determined relatively 
well because both the weak and strong vertices  have been
experimentally determined.  
The OKE and OEE graphs are expected to be the next largest 
contribution and 
model computations\cite{RMBJ92}\   show that a significant 
contribution ($\sim 30\%$) to the 
nonmesonic  decay  mode  may come from the OKE amplitude.
More importantly, it has been demonstrated that 
the ratio of neutron-induced to proton-induced 
decay widths of $\Lambda$ hypernuclei is sensitive to 
the weak OKE  amplitude \cite {benntalk95}, 
which, for the currently used values for the NNK vertices,
significantly cancels the contribution of the  OPE amplitude.
The resulting small ratio found for $^{12}_\Lambda C$ \cite {benntalk95}
is not consistent with what is seen experimentally 
\cite{BNL91,KEK92,Ejiri95},
although experimental uncertainties are large.
It also appears that vector meson exchange (e.g. $K^*, \rho, ...$)
contributes to this ratio \cite {benntalk95}.   
Such exchanges would be included in local four-baryon
$\Delta s=1$ operators in chiral perturbation theory.

It is not possible to make a direct experimental determination of
the NNK weak couplings 
that appear in the nonmesonic decay amplitudes; instead, flavour 
SU(3) is  used 
to relate these couplings to the weak pion couplings.
Previous analyses of hypernuclear decay have either ignored SU(3) 
breaking 
or assigned an arbitrary $30\%$  uncertainty 
to these couplings as an estimate of the SU(3) breaking,  eg. 
\cite{MS94} .
We will estimate the size of SU(3) breaking in the NNK amplitudes 
using 
chiral perturbation theory.

Understanding the NNK weak interactions may  shed light on a 
troubling  
situation encountered in the decay of free hyperons
outside the role they play in hypernuclear decay.
Both the S-wave and P-wave amplitudes for  $\Delta s=1$ hyperon 
decay
are well studied experimentally.
The S-wave amplitudes are adequately described, even at 
tree level, by a weak 
operator  transforming as an $(8_L,1_R)$ under chiral
${\rm SU(3)_L}\otimes  {\rm SU(3)_R}$. 
A long standing problem is that the  P-wave amplitudes are not 
well 
reproduced 
at tree level using the coupling constants extracted from the 
S-wave 
amplitudes.  A one-loop calculation of the leading SU(3) 
corrections to 
hyperon decay,
performed in ref.\cite{ej}, showed that this situation is not 
improved by
including the leading
terms nonanalytic in the strange
quark mass.  Further, these corrections
change the tree level prediction
of P-wave amplitudes by $100\%$ (typical SU(3) breaking 
corrections are  $\sim 30\%$), causing concern \cite{BSW85a,georgi} 
that chiral perturbation theory may not be valid for 
such processes.
It was suggested in ref. \cite{ej} that the problem  may instead 
be that the 
weak coupling constants  extracted from S-wave fits lead to  
(accidental) 
cancellations between the tree level  P-wave amplitudes.  
Large  SU(3) breaking effects are then a result of small 
tree level 
amplitudes,
and not a breakdown of chiral perturbation theory.  
Since there is
only one graph that contributes to P-wave NNK interactions at 
tree level, such accidental cancellations are absent.
Experimental determination of these weak NNK vertices would 
provide
an indication of the applicability of chiral perturbation theory 
to  such processes.

\section{The Chiral Lagrangian for  Nonleptonic Interactions}

At the momentum transfers characteristic of nonmesonic 
hypernuclear decay, $p < \Lambda_\chi$, the relevant degrees of 
freedom 
are the lowest mass octet and decuplet baryons and the 
pseudo-Goldstone bosons 
$\pi$, $K$, and $\eta$.
The low energy strong interaction of the these hadrons
is described by the Lagrange density 
\begin{eqnarray}\label{strongl}  
{\cal L}^{st} &=& i {\rm Tr} \bar B_v \left(v\cdot {\cal D} 
\right)B_v 
+ 2 D\  {\rm Tr} \bar B_v S_v^\mu \{ A_\mu, B_v \} 
+ 2 F\ {\rm Tr}  \bar B_v S_v^\mu [A_\mu, B_v] 
\nonumber \\
&&- i \bar
T_v^{\mu} (v \cdot {\cal D}) \  T_{v \mu} 
+ \Delta m \bar T_v^{\mu} T_{v \mu} 
+ {\cal C} \left(\bar T_v^{\mu} A_{\mu} B_v + \bar B_v A_{\mu} 
T_v^{\mu}\right)
\nonumber\\ 
&& + 2 {\cal H}\  \bar T_v^{\mu} S_{v \nu} A^{\nu}  T_{v \mu} 
+ {f^2 \over 8} {\rm Tr} \partial_\mu \Sigma \partial^\mu 
\Sigma^\dagger 
+ \mu {\rm Tr} \left( m_q\Sigma + m_q^\dagger\Sigma^\dagger \right) \ 
+\ \cdots \ \ \ \  ,
\end{eqnarray}
where $f$ is the meson decay constant, $m_q$ is the light quark 
mass matrix, 
 ${\cal D_\mu}= \partial_\mu+[V_\mu,  \; ]$ 
is the covariant chiral derivative
and use is made of the vector and axial vector chiral currents 
\begin{eqnarray}
V_\mu&=&{1 \over 2} (\xi\partial_\mu\xi^\dagger + 
\xi^\dagger\partial_\mu\xi) 
\nonumber \\
A_\mu&=&{i \over 2} (\xi\partial_\mu\xi^\dagger - 
\xi^\dagger\partial_\mu\xi) 
\ \ \ .
\end{eqnarray}
The dots in Eq.~\ref{strongl}\  represent higher dimension operators 
(involving 
more derivatives  and insertions of the  light quark mass matrix) 
whose contributions are suppressed by inverse powers  of 
$\Lambda_\chi$.
The octet baryon field of four-velocity $v$ is denoted by $B_v$ 
and has SU(3) 
elements
\begin{eqnarray}\label{octet}
 B_v =
\pmatrix{ {1\over\sqrt2}\Sigma_v^0 + {1\over\sqrt6}\Lambda_v &
\Sigma_v^+ & p_v\cr \Sigma_v^-& -{1\over\sqrt2}\Sigma_v^0 +
{1\over\sqrt6}\Lambda_v&n_v\cr \Xi_v^- &\Xi_v^0 &- 
{2\over\sqrt6}\Lambda_v
\cr }
\ \ \ ,
\end{eqnarray}
and the decuplet baryons appear as the elements of the (totally
symmetric) $T_v$:
\begin{eqnarray}\label{decuplet}
& & T^{111}_v  = \Delta^{++}_v, \ \ T^{112}_v = 
{1\over\sqrt{3}}\Delta^{+}_v, 
\ \ T^{122}_v = {1\over\sqrt{3}}\Delta^{0}_v, \ \ T^{222}_v = 
\Delta^{-}_v,
\nonumber\\
& & T^{113}_v = {1\over \sqrt{3}}\Sigma^{*+}_v,\ \ 
T^{123}_v  = {1\over\sqrt{6}}\Sigma^{*0}_v, \ \ 
T^{223}_v = {1\over\sqrt{3}}\Sigma^{*-}_v,
\ \ T^{133}_v = {1\over\sqrt{3}}\Xi^{*0}_v, 
\nonumber\\
& & T^{233}_v = {1\over\sqrt{3}}\Xi^{*-}_v, \ \ T^{333}_v = 
\Omega^-_v  \ \ \ .
\end{eqnarray}
The octet of pseudoscalar  pseudo-Goldstone bosons  
resulting from the spontaneous breaking of chiral symmetry appear 
in 
the $\Sigma$ field, with
\begin{eqnarray}
\Sigma  = \xi^2= {\rm exp}\left( {2 i M\over f} \right) \ \ \ ,
\end{eqnarray}
where
\begin{eqnarray}
M = 
\left(\matrix{{1\over\sqrt{6}}\eta+{1\over\sqrt{2}}\pi^0&\pi^+&K^
+\cr
\pi^-&{1\over\sqrt{6}}\eta-{1\over\sqrt{2}}\pi^0&K^0\cr
K^-&\overline{K}^0&-{2\over\sqrt{6}}\eta\cr}\right)
\ \ \ \  .
\end{eqnarray}
The strong couplings constants $F, D, {\cal C}$ and ${\cal H}$ 
have 
been determined from one-loop computations of axial matrix 
elements 
between octet baryons \cite{mj} 
and strong decays of decuplet baryons \cite{bss}.

The $\Delta s=1$ weak interactions of the pseudo-Goldstone bosons 
and  the lowest lying baryons are described, assuming octet
dominance, by the Lagrange density 
\begin{eqnarray}\label{weakl}
{\cal L}^{\Delta s=1}_v &=& 
G_Fm_\pi^2 f_\pi \Big( h_D {\rm Tr} {\overline B}_v 
\lbrace \xi^\dagger h\xi \, , B_v \rbrace \; 
+ \; 
 h_F {\rm Tr} {\overline B}_v  
{[\xi^\dagger h\xi \, , B_v ]} \; \nonumber \\ &&
+  h_C {\overline T}^\mu_v
(\xi^\dagger h\xi) T_{v \mu}  \; + \; 
 { h_\pi \over 8} {\rm Tr} \left(  h \, \partial_\mu 
\Sigma 
\partial^\mu 
\Sigma^\dagger  \right) 
\ + \ \cdots\ \ \ \Big) \ \ \ \   ,
\end{eqnarray}
where
\begin{eqnarray}
h = \left(\matrix{0&0&0\cr 0&0&1\cr 0&0&0}\right)  \ \ \ ,
\end{eqnarray}
and the constants $f_\pi$, $h_D, h_F, h_\pi$ and $h_C$ are 
determined experimentally.  
The pion decay constant is known to be $f_\pi \sim 132$ MeV. 
We have inserted factors of $ G_Fm_\pi^2 f_\pi $ in Eq.~\ref{weakl}\ 
so that the constants $h_D, h_F$, and $h_C$ are 
dimensionless and 
of order unity.
At tree level, the weak decay of the octet baryons 
gives\cite{mj,ej} 
$h_D = -0.58$ and $h_F = +1.40$,  while the weak
decay of the $\Omega^-$ gives 
$h_C \sim 1.4$.
The weak meson coupling $h_\pi$ is determine from nonleptonic 
kaon decays 
to be $h_\pi = 1.4$ MeV.  
The dots denote  higher dimension operators 
involving more derivatives and 
insertions of the light quark mass matrix.

We will determine the S-wave and P-wave amplitudes for weak
NNK interactions, including the  SU(3) violating one-loop 
corrections.
In the spirit of chiral perturbation theory  we compute the 
leading 
nonanalytic corrections dependent upon the mass of the strange 
quark, 
of the form $m_s\ln m_s$.
This requires the computation of one-loop graphs involving 
kaons, pions, and etas with
octet and/or decuplet baryons.   Such graphs are divergent and 
regularized in 
$n$-dimensions with modified minimal subtraction, 
$\overline{MS}$.
The divergences are absorbed by higher dimension operators whose 
coefficients 
depend upon the renormalization scale. 
The sum of the counterterm and the loop graph is scale 
independent.
By choosing to renormalize at the chiral symmetry breaking scale,
and using the fact  that the 
coefficients of the higher dimension operators are analytic 
functions of the 
light 
quarks masses,
the size of these coefficients can be {\it estimated} using naive 
dimensional 
analysis.
In the chiral limit the contributions from the higher dimension 
operators are 
subdominant compared to the logarithms that arise from the loop 
graphs
involving the lowest dimension operators.
It is these chiral logarithms  that we compute in this work.
For physical values of the kaon, pion, and eta masses, these 
quantities 
represent only 
an estimate of the size of 
SU(3) breaking effects; contributions from local counterterms 
will be 
of the same order.   Unfortunately, the coefficients of the
counterterms are not directly computable from the 
chiral Lagrangian and must be determined experimentally.

The amplitude for the weak $\Delta s=1$ NNK interactions has the 
form
\begin{eqnarray}\label{spamp}
{\cal A} = i G_F m_\pi^2\  {f_\pi \over f_K}\ \overline{N}_v \  \left[   
{\cal A}^{(S)}  + 
 2 {k \cdot S_v \over \Lambda_\chi} {\cal A}^{(P)}  \right] N_v
\end{eqnarray}
where $N_v$ contains the nucleon doublet
\begin{eqnarray}
N_v = \left(\matrix{ p_v\cr n_v }\right) \, \, \, ,
\end{eqnarray}
and $k$ is the outgoing momentum of the kaon.
The amplitudes $ {\cal A}^{(S)}$ and ${\cal A}^{(P)} $ are the 
S-wave and 
P-wave amplitudes 
respectively, and are computed below.
We have chosen to normalize ${\cal A}^{(P)} $ to $\Lambda_\chi$ 
so that it is 
dimensionless.
An explicit factor of $f_\pi/f_K$ appears in Eq.~\ref{spamp}\ and further 
we will 
distinguish $f_K$ from $f_\pi$ in the expressions arising from the 
one loop amplitudes.
There is evidence from other loop computations that this SU(3) 
breaking difference 
should be included explicitly \cite{jlms93}.

\section{Computation of Amplitudes}

There are three vertices that occur in $\Delta s=1$ weak 
nonleptonic 
interactions involving nucleons and kaons: 
$p \bar p K^0$, $n \bar p K^+$, and $n \bar n K^0$.  
They are not independent in the limit of isospin symmetry 
and are related by
\begin{eqnarray}
{\cal A}^{(L)}(nnK)  - {\cal A}^{(L)}(ppK) = {\cal A}^{(L)}(npK) 
\ \ \ ,
\end{eqnarray}
where $L=0$ (S-wave) or $L=1$ (P-wave);
this relation is true for both S-wave and P-wave amplitudes 
independently.
The amplitudes 
\begin{eqnarray}
{\cal A}^{(L)} = {\cal A}_0^{(L)} + {\cal A}_1^{(L)} + \cdots \ ,
\end{eqnarray}
where the subscript denotes the order in chiral perturbation 
theory and the 
dots indicate contributions arising from more insertions of the 
light quark mass matrix or involving more derivatives.

\subsection{S-Wave Amplitudes}

At tree level (see Fig.~\ref{tree}) the S-wave amplitudes 
appear directly from the 
first and second 
terms in the 
weak Lagrangian of Eq.~\ref{weakl}\ ,  
\begin{eqnarray}\label{stree}
{\cal A}_0^{(S)} (p \bar p K^0) &=& h_F-h_D \nonumber \\
{\cal A}_0^{(S)} (p \bar n K^+) &=& h_F+h_D \nonumber \\
{\cal A}_0^{(S)} (n \bar n K^0) &=& 2 h_F 
\ \ \ \ .
\end{eqnarray}
Experimental measurements of 
hyperon decays are used to find the parameters $h_D$ and $h_F$.  
Because the above amplitudes are tree level
expressions, it is appropriate to extract these parameters
using tree level predictions of the chiral
Lagrangian. This gives $h_D$ = --.58 and 
$h_F$ = 1.40\cite{ej}.  The tree level S-wave amplitudes
are then 2.0, 0.8, and 2.8, respectively.
The numbers will be modified by SU(3) breaking effects, as computed
below.

Direct computation of the loop graphs shown in 
Fig.~\ref{swaves} lead to 
S-wave amplitudes

\begin{eqnarray}
{\cal A}_1^{(S)} (p \bar p K^0) &=& 
{m_K^2  \over 16\pi^2 f_K^2}\ln\left({m_K^2 \over 
\Lambda_\chi^2}\right) 
\Big( {h_D\over 3} ( 1+13D^2-18DF-27F^2)   \nonumber  \\
& & \phantom{{m_K^2  \over 16\pi^2 f_K^2}\ln({m_K^2 \over 
\Lambda_\chi^2}) + 
}
+ {h_F\over 3} ( -7-7D^2+6DF+9F^2)  \Big)      \nonumber  \\
&& + {2 C^2 \over 9}h_C{\cal J}(\Delta m^{\Sigma^*}_N) 
- {\cal A}_0^{(S)} (p \bar p K^0) {\cal Z}_\Psi  
\ \ \ \ ,
\end{eqnarray}

\begin{eqnarray}
{\cal A}_1^{(S)} (p \bar n K^+) &=&
{m_K^2  \over 16\pi^2 f_K^2} 
\ln\left({m_K^2 \over \Lambda_\chi^2}\right) 
\Big( {h_D\over 3}(2-4D^2-24F+36F^2)      \nonumber  \\
& & \phantom{{m_K^2  \over 16\pi^2 f_K^2}\ln({m_K^2 \over 
\Lambda_\chi^2}) + 
}
+ {h_F\over 3}(2+20D^2-48DF+36F^2) \Big)         \nonumber  \\
&&  - {4 C^2 \over 9}h_C{\cal J }(\Delta m^{\Sigma^*}_N) 
 - {\cal A}_0^{(S)} (p \bar n K^+)  {\cal Z}_\Psi 
\ \ \ \ ,
\end{eqnarray}

\begin{eqnarray}
{\cal A}_1^{(S)} (n \bar n K^0) &=&
{m_K^2 \over 16\pi^2 f_K^2} \ln\left({m_K^2 \over 
\Lambda_\chi^2}\right) 
\Big( h_D( 1+3D^2-14DF+3F^2)          \nonumber \\
& & \phantom{{m_K^2  \over 16\pi^2 f_K^2}\ln({m_K^2 \over 
\Lambda_\chi^2}) + 
}
+ {h_F\over 3} (-5 +13D^2 -42DF + 45F^2)  \Big)              
\nonumber   \\
&& - {2 \over 9} C^2h_C{\cal J}(\Delta m^{\Sigma^*}_N)
 - {\cal A}_0^{(S)} (n \bar n K^0) {\cal Z}_\Psi
\ \ \ \ ,
\end{eqnarray}

\noindent where the contribution from wavefunction 
renormalization is
given by 

\begin{eqnarray}
{\cal Z}_\Psi  =
{m_K^2  \over 16\pi^2 f_K^2} \ln ({m_K^2 \over \Lambda_\chi^2}) 
\Big( 15F^2 - 10FD +  {17 \over 3}D^2\Big) 
+  C^2  {\cal J}(\Delta m^{\Sigma^*}_N) 
\ \ \ \ \ ,
\end{eqnarray}

\noindent and the function ${\cal J}$ is

\begin{eqnarray}
{\cal J}(\delta)  =  {1 \over 16\pi^2 f_K^2}
&&\left[(m_K^2  -2 \delta^2)\ln\left({m_K^2 \over 
\Lambda_\chi^2}\right) 
+ 2\delta \sqrt{\delta^2-m_K^2 } 
 \ln \left( { \delta-\sqrt{\delta^2-m_K^2+ i\epsilon}  \over
 \delta+\sqrt{\delta^2-m_K^2 + i\epsilon}}\right) \right]
\ \ .
\end{eqnarray}
\noindent For S-waves, $\delta=\Delta 
m^{\Sigma^*}_N=m_{\Sigma^*}-m_N$.
The kaon decay constant $f_K$ = 1.22 $f_\pi$.

We have not included wavefunction renormalization of the external 
meson 
field in our KNN amplitude computations since the kaons do not 
appear as 
asymptotic states in the weak scattering processes under  
consideration.

In order to determine the ${\cal A}^{(S)}$ we insert the axial 
coupling 
constants
$D, F ,{\cal C}$, and ${\cal H}$ extracted by the one-loop 
computations 
of 
\cite{mj,bss}: 
\begin{eqnarray}\label{strongFDCH}
D & = &  0.6\pm 0.1\ \ ,\ \ F = 0.4\pm 0.1\ \     \nonumber \\
{\cal C} & = & -1.2\pm 0.1\ \ ,\ \ {\cal H} = -2.0\pm 0.2
\ \ \ \ ,
\end{eqnarray}
and the weak coupling constants determined at one-loop level 
\cite{ej}
\begin{eqnarray}\label{weakFDCH}
h_D & = & -0.35 \pm 0.09 \ \ ,\ \ h_F = 0.86\pm 0.05  \nonumber 
\\
h_C & = & -0.36\pm 0.65
\ \ \ .
\end{eqnarray}
The uncertainties in these couplings are treated as uncorrelated 
for the purpose of determining the uncertainty in the NNK 
amplitudes.   
We determine the error in the NNK amplitudes by varying the
parameters over their 
allowed range and require that the choice of parameters 
reproduce the $\Delta s=1$ S-wave hyperon amplitudes within their 
uncertainties.
In this way we determine that, with $\Lambda_\chi = 1$ GeV,

\begin{eqnarray}
{\cal A}_0^{(S)} (p \bar p K^0)  +  {\cal A}_1^{(S)} (p \bar p 
K^0) & = &  
   1.5\pm 0.1 \nonumber \\ 
{\cal A}_0^{(S)} (p \bar n K^+)  +  {\cal A}_1^{(S)} (p \bar n 
K^+) & = &  
   0.4\pm 0.1  \nonumber \\ 
{\cal A}_0^{(S)} (n \bar n K^0)  +  {\cal A}_1^{(S)} (n \bar n 
K^0) & = &  
   1.9\pm 0.1 
\end{eqnarray}

The SU(3) corrections tend to suppress the couplings compared to 
their 
tree level values, 
a significant contribution of which comes from the use of the 
one-loop 
extracted weak 
couplings in the tree level amplitudes instead of those extracted 
at 
tree level.
The corrections to the $ppK$ and $nnK$ couplings are, as one 
expects for 
SU(3) breaking,  
at the $30\%$ level.
However, the corrections to the $pnK$ amplitude lead to an 
effective coupling 
approximately half the  strength computed at tree level.

\subsection{P-Waves}

The tree level P-wave amplitudes come directly from 
pole graphs involving one weak vertex from Eq.~\ref{weakl} and 
one strong vertex from Eq.~\ref{strongl}, (see Fig.~\ref{tree})
\begin{eqnarray}\label{ptree}
{{\cal A}_0^{(P)} (p \bar p K^0) \over \Lambda_\chi} &=& 
   -{(D-F)(h_D-h_F) 
\over m_N-m_\Sigma}
                 \nonumber \\
{{\cal A}_0^{(P)} (p \bar n K^+) \over \Lambda_\chi} &=& -{1 \over 
6}{(D+3F)(h_D+3h_F) 
                 \over m_N-m_\Lambda} + {1 \over 
2}{(D-F)(h_D-h_F) 
                 \over m_N-m_\Sigma}\nonumber \\
{{\cal A}_0^{(P)} (n \bar n K^0) \over \Lambda_\chi} &=& -{1 \over 
6}{(D+3F)(h_D+3h_F) 
                 \over m_N-m_\Lambda}-{1 \over 2}{(D-F)(h_D-h_F) 
                 \over m_N-m_\Sigma}
\ \ \ \ .
\end{eqnarray}

\noindent The numerical values for these amplitudes
are found by using the parameters (extracted
from tree level comparison to experiment)
$h_D$ = --0.58, $h_F$ = 1.40, $D$ = 0.8, and $F$ = 0.5 \cite{ej,man}.  
This yields tree level P-amplitudes of  --2.4, 9.1, and 6.7 
$(\times \Lambda_\chi / 1 \, {\rm GeV})$, respectively.  
Loop diagrams as shown in Fig.~\ref{pwaves} give

\begin{eqnarray}\label{plooppp}
{{\cal A}_1^{(P)} (p \bar p K^0) \over \Lambda_\chi} &=&  
{m_K^2  \over 16\pi^2 f_K^2} \ln\left({m_K^2 \over 
\Lambda_\chi^2}\right) \times \nonumber\\
& & \left[
{ h_D \over m_N-m_\Sigma}  
\left[ {10\over 3}D - {10\over 3}F + {92\over 9}D^3 - {140\over 
9}D^2F
      +{20\over 3}DF^2 - 4 F^3 \right] \right. \nonumber\\
& &   \left. 
 +  { h_F \over m_N-m_\Sigma}
\left[ {10\over 3}F - {10\over 3}D  - {56\over 9}  D^3 + 
{140\over 9}D^2F
      - {32\over 3}DF^2 + 4 F^3 \right]
 \right.\nonumber\\
& & \left. 
+ h_\pi \left[ {5 \over 36} D + {7 \over 36}F - {41 \over 36}D^3
               - {65 \over 36}D^2F + {313 \over 36}DF^2 - 
    {37 \over 12}F^3 - {5 \over 6}C^2(D-F)\right]
\right]
\nonumber\\
&-& {10 \over 81} C^2 {\cal H} {h_D-h_F\over m_N-m_\Sigma} 
{\cal J}\big(\Delta m^{\Sigma^*}_N\big)
- {2 \over 9} {D-F \over m_N-m_\Sigma} C^2 h_C
{\cal J}\big(\Delta m^{\Sigma^*}_N\big) \nonumber \\
&-&{2 \over 9} C^2{h_D-h_F \over m_N - m_\Sigma}
     \big[ (D + 5F) {\cal K}(m_K,\Delta m^{\Sigma^*}_N)
-(D-3F){\cal K}(m_\eta,\Delta m^{\Sigma^*}_N) \big] \nonumber \\ 
&+& {10 \over 81}h_\pi C^2 {\cal H} {\cal J}(\Delta m^{\Delta}_N) 
\nonumber \\
&+& h_\pi C^2\Big[{4 \over 9}(D-F){\cal K}(m_K,\Delta m^\Delta_N)
    +{2 \over 9}(F+D){\cal K}(m_K,\Delta m^{\Sigma^*}_N) 
\nonumber \\
& & \phantom{ h_\pi C^2\Big[} 
-{1\over 27}(D-3F)G_{\eta K\Sigma^*} - {1 \over 12}(D-F)
 \tilde{G}_{\eta K \Sigma^*} \Big] 
\nonumber \\
&+& {(D-F)(h_D-h_F) \over m_N-m_\Sigma} {14\over 3}C^2{\cal 
J}(\Delta 
m_\Sigma)
-{\cal A}_0^{(P)}(p \bar p K^0){\cal Z}_\Psi 
\ \ \ ,
\end{eqnarray}

\noindent where ${\cal Z}_\Psi$ is the same wavefunction 
renormalization
employed for the S-wave expressions.

\begin{eqnarray}\label{plooppn}
{{\cal A}_1^{(P)} (p \bar n K^+) \over \Lambda_\chi} &=&
{m_K^2  \over 16\pi^2 f_K^2} \ln ({m_K^2 \over \Lambda_\chi^2}) 
 \Big({h_D \over m_N-m_\Lambda}
     \Big[{5\over 9}D + {5\over 3}F + {2\over 27}D^3 + 2 D^2F 
	  + {34 \over 3}DF^2 + 6F^3\Big] \nonumber \\
&& \hskip 1.1cm + {h_F \over m_N-m_\Lambda}
      \Big[{5\over 3}D + 5F + {20 \over 9}D^3 + 6D^2F
            + 16 DF^2 + 18 F^3\Big] \nonumber \\
&& \hskip 1.1cm + {h_D \over m_N-m_\Sigma}
       \Big[ - {5\over 3}D + {5\over 3}F - {46 \over 9}D^3 + {70 
\over 9}D^2F
          - {10\over 3} DF^2 + 2 F^3\Big] \nonumber \\
&& \hskip 1.1cm + {h_F \over m_N-m_\Sigma}
       \Big[{5\over 3}D - {5\over 3}F + {28 \over 9}D^3 - {70 
\over 9}D^2F
                      + {16 \over 3}DF^2 - 2  F^3\Big] \nonumber 
\\
& & \hskip 1.1cm - h_\pi  \Big[{11 \over 36}D + {11 \over 36}F  - 
{101 \over 108}D^3- {23\over 12}D^2F + {23\over 36}DF^2 - 
{17\over 12}F^3 + C^2({1 \over 24} F-{13 \over 24} D)
 \Big] \Big) \nonumber\\
& + & {5 \over 9}C^2{\cal H} {\cal J}(\Delta m_N^{\Sigma^*}) 
\Big[{1 \over 3}{h_D+3h_F \over m_N-m_\Lambda}+
                 {1 \over 9}{h_D-h_F\over m_N-m_\Sigma}\Big] 
\nonumber \\
&+&C^2\Big[\Big(-{1 \over 3}{(h_D+3h_F)(D+F) \over 
m_N-m_\Lambda}+
			{1 \over 9} {(h_D-h_F)(D+5F) \over 
m_N-m_\Sigma}\Big)
                 {\cal K}(m_K,\Delta m_N^{\Sigma^*}) \nonumber \\
        && \hskip 1.1 cm-{1 \over 9}{(h_D-h_F)(D-3F) \over 
m_N-m_\Sigma}
        {\cal K}(m_\eta,\Delta m_N^{\Sigma^*})\Big] \nonumber \\                        
&+& C^2 h_C {\cal J}(\Delta m_N^{\Sigma^*})
\Big[{1 \over 9}{D-F \over m_N-m_\Sigma} + {1 \over 3}
      {D+3F \over m_N-m_\Lambda}\Big] 
+  {10 \over 81} h_\pi C^2 {\cal H} {\cal J}(\Delta 
m_N^{\Sigma^*}) 
\nonumber \\ 
&+&h_\pi C^2\Big[{4 \over 9}(D-F){\cal K}(m_K,\Delta m_N^\Delta)-
          {5 \over 18}(F+D){\cal K}(m_K,\Delta m_N^{\Sigma^*}) 
\nonumber \\
& & \phantom{ h_\pi C^2\Big[}
-{1\over 54}(D-3F)G_{\eta K\Sigma^*}+ 
{1 \over 24}(D-F) \tilde{G}_{\eta K \Sigma^*} \Big] 
\nonumber \\
&+& C^2 \Big[{1 \over 3}{(h_D+3h_F) \over m_N-m_\Lambda}(D+3F)
-{7 \over 3} {h_D-h_F \over m_N-m_\Sigma}(D-F)  \Big]
         {\cal J}(\Delta m_\Sigma) \nonumber \\
&-&{\cal A}_0^{(P)}(n \bar p K^+){\cal Z}_\Psi
\ \ \ \ ,
\end{eqnarray}

\begin{eqnarray}\label{ploopnn}
{{\cal A}_1^{(P)} (n \bar n K^0) \over \Lambda_\chi} &=&
{m_K^2  \over 16\pi^2 f_K^2} \ln ({m_K^2 \over \Lambda_\chi^2})
\Big({h_D \over m_N-m_\Lambda}
		\Big[{5\over 9}D + {5\over 3}F + {2 \over 27}D^3 
+ 2 D^2F
                      + {34 \over 3}DF^2 + 6 F^3\Big] \nonumber 
\\
&& \hskip 1.1cm + {h_F \over m_N-m_\Lambda}
                    \Big[ {5\over 3}D + 5 F + {20\over 9}D^3 + 
6D^2F
                         + 16 DF^2 + 18 F^3\Big] \nonumber \\
&& \hskip 1.1cm + {h_D \over m_N-m_\Sigma}
		    \Big[{5\over 3}D - {5\over 3}F + {46 \over 9} 
D^3 
			- {70\over 9}D^2 F + {10 \over 3}DF^2 - 2 
F^3\Big] 
\nonumber \\
&& \hskip 1.1cm +{h_F \over m_N-m_\Sigma}
		\Big[-{5\over 3}D + {5\over 3}F - {28 \over 9}D^3 
+ {70 \over 
9}D^2F
  -{16 \over 3}DF^2 + 2 F^3\Big] 
\nonumber \\
&& \hskip 1.1 cm -  h_\pi \Big[{1 \over 6} D + {1 \over 9}F 
	+{11\over 54}D^3 - {1\over 9} D^2F - {145 \over 18} DF^2 
+ {5\over 3}F^3 + C^2({7 \over 24}D-{19\over 24}F)\Big]\Big)
\nonumber\\
&+& {5  \over 9}C^2{\cal H} {\cal J}(\Delta m_N^{\Sigma^*})
\Big[{1 \over 3}{h_D+3h_F\over m_N-m_\Lambda}-
                   {1 \over 9}{h_D-h_F\over m_N-m_\Sigma}\Big] 
\nonumber \\
&+&C^2\Big[\Big(-{1 \over 9}{(h_D-h_F)(D+5F) \over m_N-m_\Sigma}
                 -{1 \over 3}{(h_D+3h_F)(D+F) \over m_N - 
m_\Lambda}\Big)
 {\cal K}(m_K,\Delta m_N^{\Sigma^*}) \nonumber \\
&& \hskip 1.1cm +\Big({1 \over 9}{(h_D-h_F)(D-3F) \over 
m_N-m_\Sigma}\Big)
{\cal K}(m_\eta,\Delta m_N^{\Sigma^*})\Big] \nonumber \\
&-& C^2 h_C {\cal J}(\Delta m_N^{\Sigma^*})
   \Big[{1 \over 9}{D-F\over m_N-m_\Sigma}
        -{1 \over 3}{D+3F\over m_N-m_\Lambda}\Big] 
+ {20 \over 81}h_\pi C^2{\cal H} {\cal J}(\Delta m_N^{\Delta})
\nonumber \\
&+& h_\pi C^2 \Big[
    {8 \over 9}(D-F){\cal K}(m_K,\Delta_N^\Delta)-
                   {1 \over 18}(F+D){\cal K}
(m_K,\Delta_N^{\Sigma^*})
\nonumber \\
& & \phantom{ h_\pi C^2 \Big[} 
 -{1\over 54}(D-3F) G_{\eta K \Sigma^*}-{1 \over 24}(D-F)
\tilde{G}_{\eta K \Sigma^*} \Big]
\nonumber \\
&+& C^2  \Big[{1 \over 3}{h_D+3h_F \over m_N-m_\Lambda}(D+3F)+
{7 \over 3}{h_D-h_F\over m_N-m_\Sigma}(D-F)\Big]{\cal J}(\Delta 
m_\Sigma)
 \nonumber \\
&-&{\cal A}_0^{(P)}(n \bar n K^0){\cal Z}_\Psi  
\ \ \ \ .
\end{eqnarray}

\noindent The function ${\cal K}(m,\Delta)$ which appears in 
Eq.~\ref{plooppp}, 
Eq.~\ref{plooppn} and Eq.~\ref{ploopnn} 
from  diagrams having both decuplet and octet intermediate states 
is 
\begin{eqnarray}
{\cal K}(m,\delta)  =  {1 \over 16\pi^2 f_K^2} & & \Big\{
(m^2 - {2\over 3}\delta^2) \ln({m^2 \over \Lambda_\chi^2}) 
\nonumber\\
& & +{2 \over 3}{1 \over \delta}\Big[(\delta^2-m^2)^{3/2}
   \ln\left( {\delta-\sqrt{\delta^2-m^2+i\epsilon} \over 
       \delta+\sqrt{\delta^2-m^2+i\epsilon}}\right)
+ \pi m^3 \Big]\Big\}
\ \ \ ,
\end{eqnarray}

\noindent and the functions $G_{m_1,m_2,B}$ and
$\tilde{G}_{m_1,m_2,B}$ are given by 
\begin{eqnarray}
G_{m_1,m_2,B} & = & {m_1^2 \over m_1^2-m_2^2}
    {\cal K}(m_1,\Delta_N^B) + {m_2^2 \over m_2^2-m_1^2}
    {\cal K}(m_2,\Delta_N^B)\ \  , \nonumber \\
\tilde{G}_{m_1,m_2,B} & = & {m_1^2 \over m_1^2-m_2^2}
    {\cal J}(m_1,\Delta_N^B) + {m_2^2 \over m_2^2-m_1^2}
    {\cal J}(m_2,\Delta_N^B)\ \ \ \ .
\end{eqnarray}

\noindent  The mass differences that appear in Eq.~\ref{plooppp}, 
Eq.~\ref{plooppn} 
and 
Eq.~\ref{ploopnn}\ are  defined by 
\begin{eqnarray}
\Delta m^{\Sigma^*}_N & = &  m_\Sigma^* - m_N \ \ ,\ \  
\Delta m^\Delta_N  = m_\Delta-m_N  \nonumber\\
\Delta m_\Sigma & = & m_{\Xi^*} - m_\Sigma \ \sim \  \Delta 
m^{\Sigma^*}_N
 \ \ \  .
\end{eqnarray}

In the same way that the S-wave KNN amplitudes and associated 
uncertainties
were determined, we use the expressions for S-wave hyperon decay 
\cite{ej},
along with experimental measurements, to generate P-wave KNN 
amplitudes 
consistent 
with S-wave hyperon decay rates.
The results, with $\Lambda_\chi=1$ GeV, are
\begin{eqnarray}
{\cal A}_0^{(P)} (p \bar p K^+)  +  {\cal A}_1^{(P)} (p \bar p 
K^+) & = & 
    -1.7\pm 0.2  \nonumber \\
{\cal A}_0^{(P)} (p \bar n K^+)  +  {\cal A}_1^{(P)} (p \bar n 
K^+) & = &  
    7\pm 1   \nonumber \\
{\cal A}_0^{(P)} (n \bar n K^+)  +  {\cal A}_1^{(P)} (n \bar n 
K^+) & = &  
    6\pm 1    \nonumber \\
\end{eqnarray}

The P-wave amplitudes are seen to be reduced by $\sim 30\%$ from 
their tree level values
by the SU(3) breaking 
one-loop contributions.  This is in contrast to the $\Delta s=1$ 
hyperon decay P-wave 
amplitudes, where the corrections are at the $100\%$ level.
The NNK SU(3) breaking is the size one would naively guess  and is 
consistent with the idea
\cite{ej} that the large corrections to the P-wave  amplitudes 
for $\Delta s=1$
hyperon decays are the result of accidentally small tree level 
amplitudes, 
and not a breakdown of chiral perturbation theory.
Table I summarizes our findings. \footnote{We thank C. Bennhold for pointing
out a factor of 2 error in the original numerical values given for
the P-waves in the table.  This is corrected in the table shown.}

\section{Discussion}

Weak NNK amplitudes that contribute to nonmesonic hypernuclear 
decay 
are not directly measurable but can be related to $\Delta s=1$ 
hyperon 
decay by flavour SU(3).
We have computed the leading SU(3) breaking contributions to 
these amplitudes
using heavy baryon chiral perturbation theory and find that such
corrections can suppress both the S-wave and P-wave amplitudes 
by up to  $50\%$.
\begin{table}
\begin{tabular}{ccccc} 
& \multicolumn{2}{c}{\em S-waves} & \multicolumn{2}{c}{\em P-waves} \\ 
{\em vertex} 
& ${\cal A}^{(S)}_0$ &  ${\cal A}^{(S)}_0+{\cal A}^{(S)}_1 $
& ${\cal A}^{(P)}_0$ &  ${\cal A}^{(P)}_0 + {\cal A}^{(P)}_1$ 
\\   \tableline
\rule{0cm}{0.5cm}
$p{\overline p}K^0$  	
& 2.0	 
&$ 1.5\pm 0.1 $	 
&$-2.4$	 
&$-1.7\pm 0.2 $	 
\\  \tableline
\rule{0cm}{0.5cm}
$p{\overline n}K^+$ 
&0.8	 
&$ 0.4\pm 0.1 $	 
&9.1	 
&$ 7\pm 1 $	 
\\  \tableline
\rule{0cm}{0.5cm}
$n{\overline n}K^0$ 
&2.8	 
&$ 1.9\pm 0.4 $	 
&6.7	 
&$ 6\pm 1 $	 
\\
\end{tabular}
\vskip 0.5cm
\caption{The S-wave and P-wave amplitudes at tree level and at
one-loop. We have set $\Lambda_\chi = 1$ GeV in both the S-wave and 
P-wave amplitudes.}
\end{table}
In $\Delta s=1$ mesonic hyperon decay there are two tree level 
graphs 
contributing to P-wave amplitudes which tend to cancel against 
each other
for the values of weak 
couplings constants determined from  S-wave hyperon decay
amplitudes.
Since there is only one graph contributing to the P-wave NNK 
vertices, no 
such 
cancellations arise and the amplitudes are, in general, less 
susceptible 
to large SU(3) violation. 
It would be interesting to compare the P-wave amplitudes 
extracted from 
hypernuclear decay with the amplitudes computed in this work.
It would help us to determine if the disagreement between the 
observed 
and predicted
$\Delta s=1$ mesonic hyperon decay P-wave amplitudes is an 
accident of 
nature or a hint that chiral perturbation theory is not 
applicable to  these
processes.

We stress that our computation is only an estimate of SU(3) 
breaking effects 
as  there are unknown counterterms that also contribute.
The computation that we have performed is the leading effect in 
the chiral 
limit,  $m_q\rightarrow 0$.
There is no reason to suspect that the counterterms cancel the 
loop  contributions since 
the counterterms arise from UV physics whereas the nonanalytic 
terms from the  loop  graphs are IR effects.

In order to determine the impact of our work on the understanding 
of  hypernuclear  decay
the NNK amplitudes, including the SU(3) breaking 
corrections, 
must be incorporated into a realistic hypernucleus in the same 
way that 
previous estimates of interaction strengths have been included, 
e.g.  \cite{RMBJ92}.
It seems likely that the results found in this work will have 
significant impact on theoretical predictions for the ratio
of neutron-induced to proton-induced decay widths of 
$\Lambda$-hypernuclei, eg. $^{12}_\Lambda C$ \cite{benntalk95},
and possibly other flavours of hypernuclei.

\section{Acknowledgements}

We would like to thank the Institute for Nuclear Theory at the 
University of
Washington for their kind hospitality during some of this work. 
RPS would also like to thank the Institute for Theoretical 
Physics at the
University of California, Santa Barbara, and Carnegie-Mellon
University, where some of this work was completed.
MJS would like to thank Duke University where some of this work 
was carried out. 
MJS would also like to thank L. Kisslinger  and C. Bennhold for
useful discussions.
Our work is supported in part by the US Dept. of Energy under 
grant number 
DE-FG02-91-ER40682, and grant number DE-FG05-90ER40592.

\begin{figure}
\epsfxsize=10cm
\hfil\epsfbox{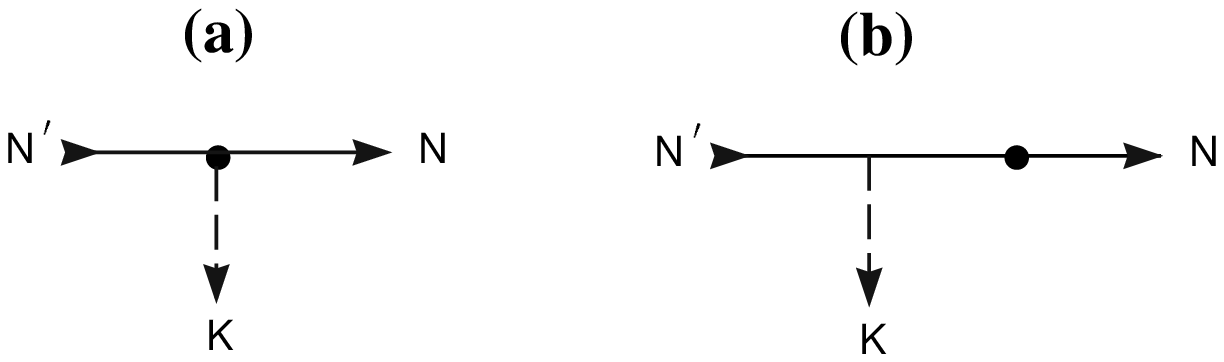}\hfill
\caption{Tree level diagrams for the KNN amplitudes.  (a) is the
S-wave diagram, and (b) is the P-wave diagram. The dashed lines
are mesons, and the solid lines are octet baryons.  An unmarked
vertex represents a strong interaction and the black dots are weak 
vertices.}
\label{tree}
\end{figure}

\begin{figure}
\epsfxsize=10cm
\hfil\epsfbox{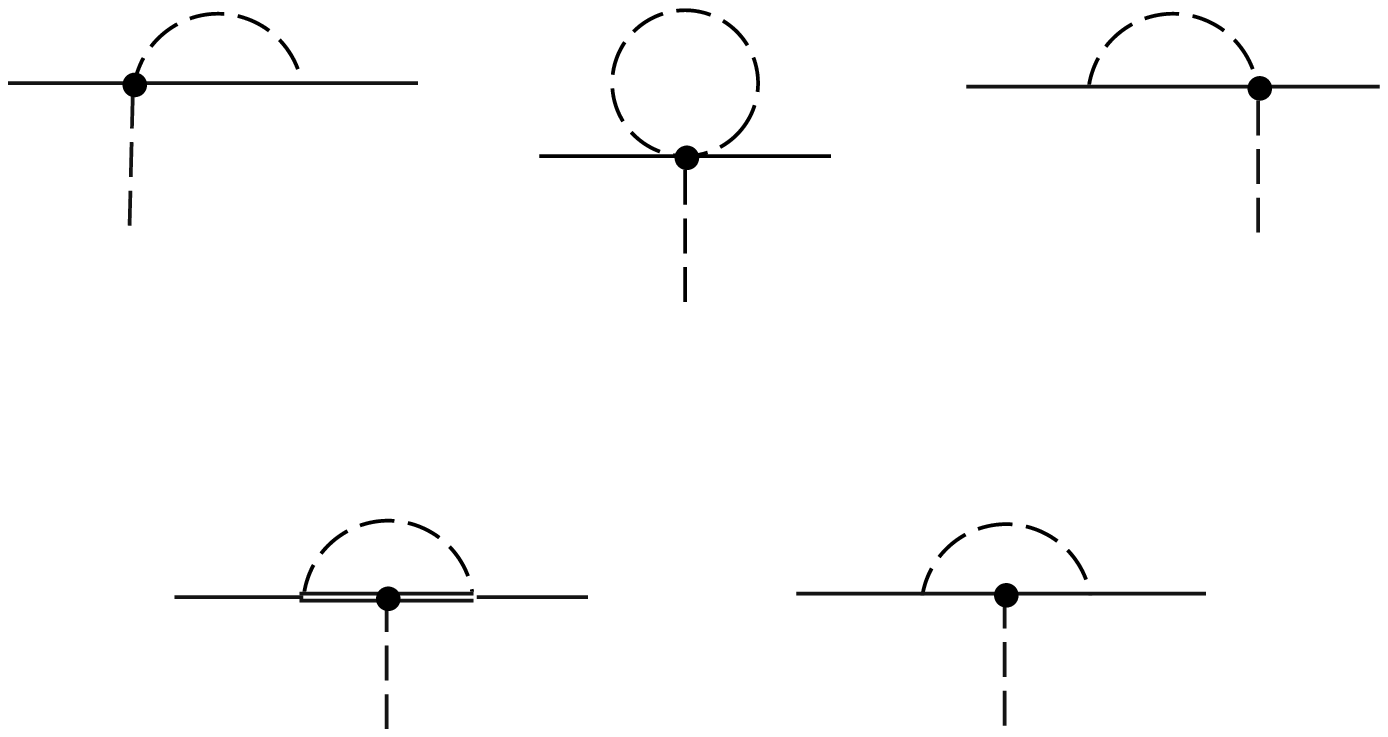}\hfill
\caption{One-loop graphs contributing to S-wave KNN 
amplitudes.  Dashed lines are mesons, solid lines are octet baryons,
and the double line indicates a decuplet baryon.
An unmarked
vertex represents a strong interaction and the black dots are weak 
vertices. The wavefunction renormalization graphs are not shown.}
\label{swaves}
\end{figure}

\begin{figure}
\epsfxsize=10cm
\hfil\epsfbox{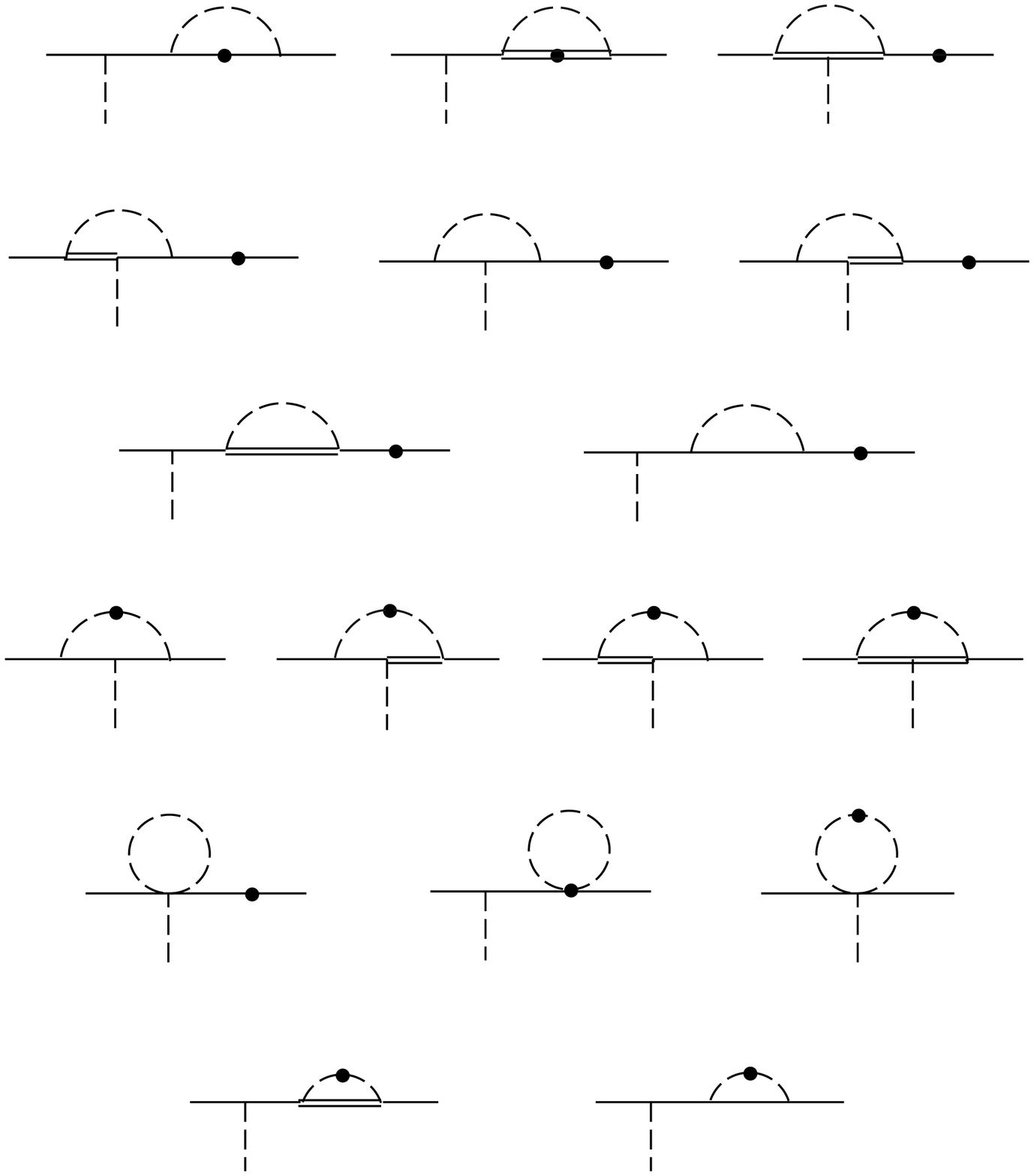}\hfill
\caption{One-loop graphs contributing to P-wave KNN 
amplitudes. Dashed lines are mesons, solid lines are octet baryons,
and the double line indicates a decuplet baryon.
An unmarked
vertex represents a strong interaction and the black dots are weak 
vertices. The diagonal wavefunction renormalization
graphs are not shown.}
\label{pwaves}
\end{figure}

\end{document}